# SUBARU prime focus spectrograph: integration, testing and performance for the first spectrograph

F. Madec *[a], A. Le Fur [a], D. Le Mignant [a], K. Dohlen [a], R. Barrette [a], M. Belhadi [a], S. Pascal [a], S. Smee [b], J. Gunn [d], J. Le Merrer [a], M. Lorred [a], M. Jaquet [a], P. Balard [a], P. Blanchard [a], W. Thao [a], F. Roman [a], V. Lapere [f], JF Gabriel [f], C. Loomis [d], M. Golebiowski [b], M. Hart [b], L. Oliveira [e], A. Oliveira [e], N. Tamura [c], A. Shimono [c]

[a] Aix Marseille Université - CNRS, LAM (Laboratoire d'Astrophysique de Marseille), UMR 7326, 13388, Marseille, France
[b] Department of Physics and Astronomy, Johns Hopkins University, Baltimore, MD, USA 21218;
[c] Kalvi Institute for the Physics and Mathematics of the Universe (WPI), University of Tokyo, Japan
[d] Princeton University, Princeton, NJ, USA 08544
[e] MCT/LNA –Laboratório Nacional de Astrofísica, Itajubá - MG - Brazil

## ABSTRACT

The Prime Focus Spectrograph (PFS) of the Subaru Measurement of Images and Redshifts (SuMIRe) project for Subaru telescope consists in four identical spectrographs fed by 600 fibers each. Each spectrograph is composed by an optical entrance unit that creates a collimated beam and distributes the light to three channels, two visibles and one near infrared. This paper presents the on-going effort for the tests & integration process for the first spectrograph channel: we have developed a detailed Assembly Integration and Test (AIT) plan, as well as the methods, detailed processes and I&T tools. We describe the tools we designed to assemble the parts and to test the performance of the spectrograph. We also report on the thermal acceptance tests we performed on the first visible camera unit. We also report on and discuss the technical difficulties that did appear during this integration phase. Finally, we detail the important logistic process that is require to transport the components from other country to Marseille.

## 1. INTRODUCTION

The Prime Focus Spectrograph (PFS) [1] of the Subaru Measurement of Images and Redshifts (SuMIRe) project for Subaru telescope is developed by a large consortium over the world lead by the Japanese of IPMU divided into two main systems, a fiber positioner placed at the prime focus of SUBARU telescope and a spectrograph composed by four identical spectrographs feed by 600 fibers each. A spectrograph module [2] covers a spectral range between 380 to 1260nm splitted in 3channels, the blue, the red and the near infrared (NIR) one. A module, as shown on Figure 1 is fed by a pseudo slit placed at the focal plane of a Schmidt collimator, the beam is split into three channels by two dichroics followed the Schmidt corrector of the collimator and the Volume Phase Holographic Gratings (VPHG). The beam, then enters into the camera, first composed by the two Schmidt correctors, followed by the Schmidt-Mangin mirror, the field lens and the focal plane composed by either two detectors for the visible camera or a single H4RG for the near infrared camera.

The spectrograph system integration is performed at LAM from dewars, focal plane assemblies and NIR cameras received from JHU[3], the fiber slit from LNA and other opto-mechanical parts from a French vendor, Winlight System.

*fabrice.madec@lam.fr  www.lam.fr

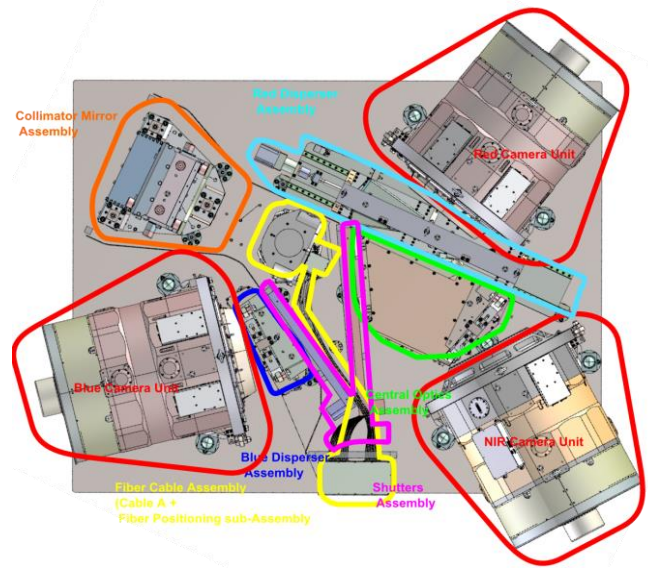

Figure 1 Overview of a Spectrograph Module

## 2. AIT PLAN

### 2.1 Plan and method

The PFS project is developed by a large consortium [1]. For the Spectrograph System (SpS), five institutes or industrial partner are involved in a hardware contribution as shown on Figure 2.

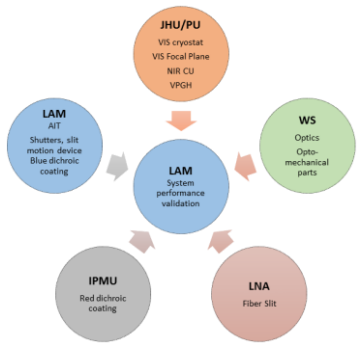

The performance validation of the SpS is performed at LAM in its 400m2 clean room facility. The assembly of the eight visible cameras is also done at LAM with some hardware components coming from JHU and Winligth System.

The number of identical parts in the PFS instrument, and the large dimensions of the optical components was a main driver to study solution and tools to minimize the integration activities. We detail the tools we have developed in section 2.2 & 2.3

One main AIT functional requirement early identified in the project was to minimize the integration activities at the summit of Mauna Kea. Consequently, we specified that the assemblies shall be repositionable without any alignment on the main bench and that assemblies shall not be dismounted nor disassembled. Due to the large size of the components and the very limited space between them, we spent particular attention on the handling and guiding tools.

Figure 2 SpS AIT partners

To develop the AIT plan, we made a flow chart that details every activity and that identifies what is needed for it. The need types are manpower, resources, tools, software. We also document every product required for any given task. The tests are identified along with the requirement they will validate. One can see on Figure 3, an example of a flow chart for the first spectral channel of a spectrograph module. Each type of activities is identified by a different color and the same logic is used in the MS project. Some information such as product name, delivered by, the requirement, are linked to excel files. We used external data visio capabilities to produce the data form.

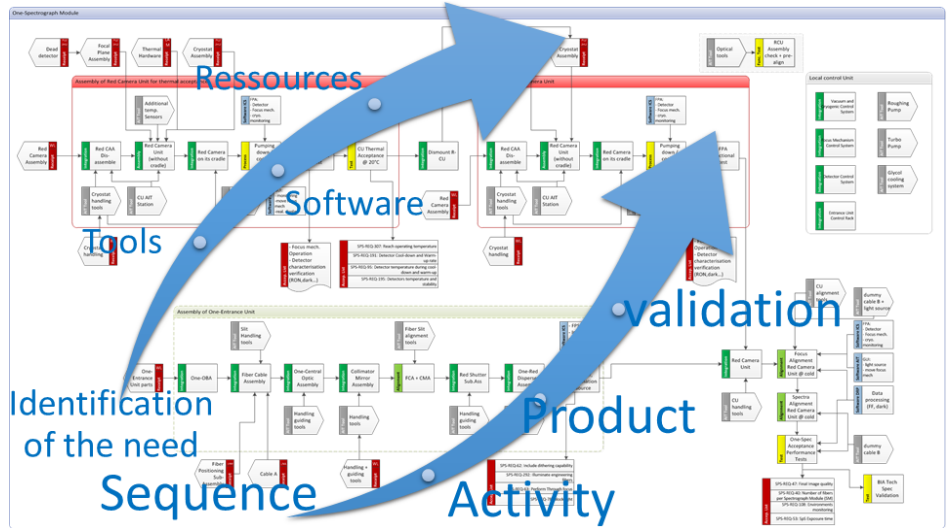

Figure 3 AIT flow chart

## 2.2 Assembly and alignment tools

### 2.2.1 Slit Alignment tool

The fiber slit needs to be aligned on the bench. The optical alignment consists in centration, focus and tilt adjustment. The instrument requires two motions: a dither and a focus adjustment. Yet, during integrations, more degrees of freedom are required. It was thus decided to place the fiber slit on an hexapod. The main advantage of using an hexapod is to easily define the rotation center as this will help us to perfectly align the slit and define the dither and focus axis correctly.

The fiber slit is curved and is 130mm heigh. We need to scan at least the central fiber and one extreme fiber to adjust its focus. This is achieved using a dedicated collimator with a mirror in the pupil plane that allows us to scan the fiber slit. The collimator has a 125mm aperture with a 1500mm focal length. The scan mirror in the pupil is also placed on an hexapod. The setup is shown in Figure 4 below.

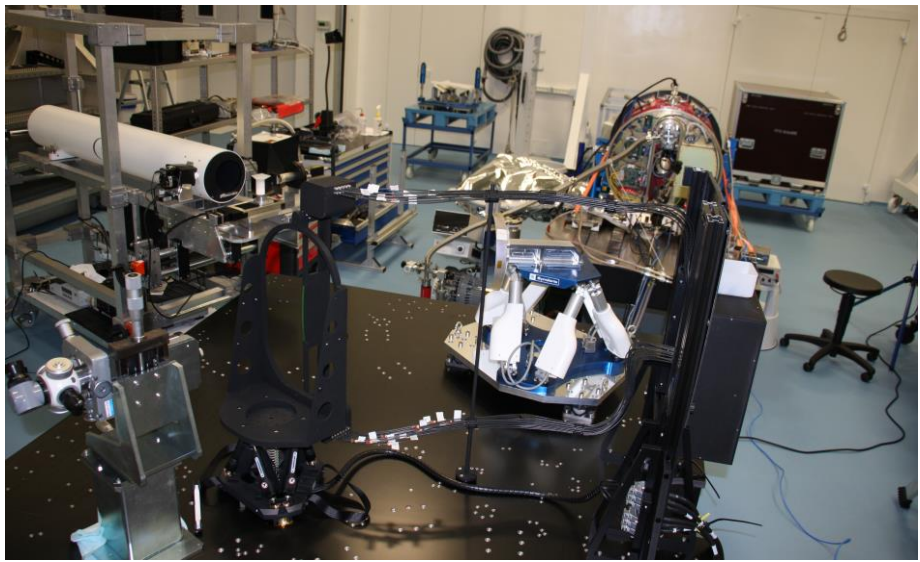

Figure 4 Fiber Slit Alignment tool

The optical alignment is performed in two steps: first the centration is done using a micro-alignment telescope (MAT). The optical reference on the bench is given by a dedicated tool used to align all the other components. This reference tool has a window and a reticule that define the optical axis of the spectrograph as well as two other reticules that define the extreme fiber positions. So we first align the MAT, the centration is done wrt the reticule and angles are tuned by auto-collimation on the window. Then using a vertical translation stage, we will seek the two crosshair reticules and write their positions. The scanning mirror and the collimator are then aligned using the MAT on its central position. In a second phase, we place the fiber slit on the bench and do centration using the MAT. We will use the MAT to adjust the best possible each degrees of freedom of the slit.

The fiber slit has two type of fibers: engineering and science fibers. The engineering fibers have a smaller diameter (80µm) than the science fiber (125µm). The engineering fibers are placed along the slit to a set of selected positions to scan the field and wavelength parameter spaces. In addition, we put several fibers around the slit center as well as the two extreme positions of the fibers.

Almost all processes are to be automated and we have designed the alignment tools so that they be remotely controlled. A dedicated gui shown on Figure 5 Slit alignment software panelFigure 5 was developed to perform this alignment. We control each component: the collimator camera and its translation stage, the scanning mirror, the sources and the slit stage. We also have in real-time the display of the image and the result of the through-focus analysis.

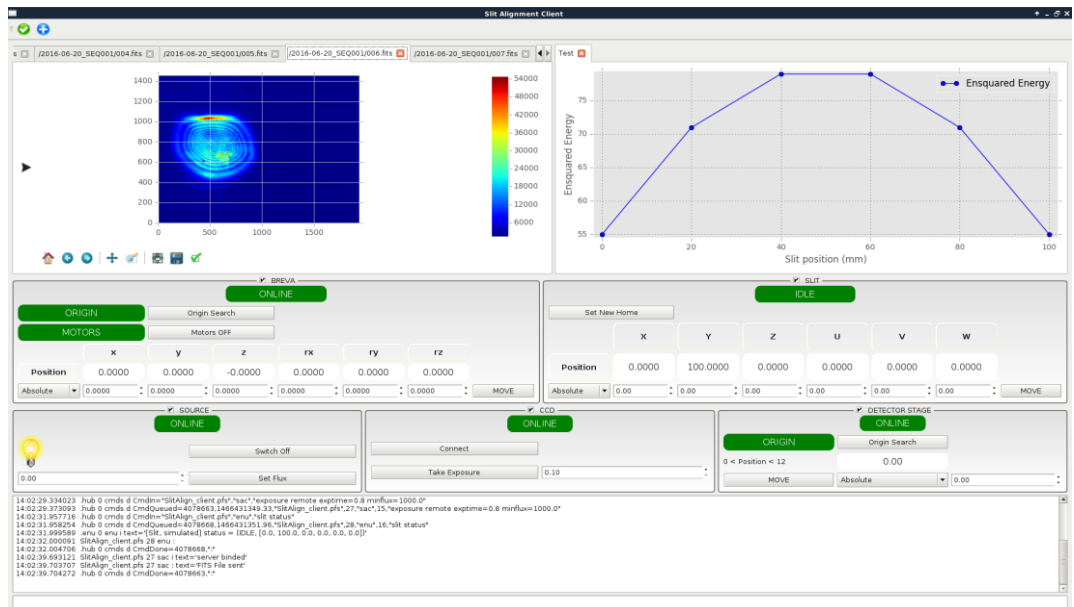

Figure 5 Slit alignment software panel

### 2.2.2 CU AIT Station

The main specification of the CU AIT Station is to allow the assembly of the visible camera unit without losing the optical alignment of the Schmidt camera. There are two main functions: the first one is the integration of the focal plane assembly[4][6] inside the camera and the second one is to close the cryostat and finalize the assembly of the camera unit.

The camera assembly is received from Winlight System with the optics aligned. The cryostat vessel and the focal plane [5] is delivered by JHU/PU. LAM has to integrate the focal plane inside the Schmidt camera and install the cryostat vessel.

The CU AIT Station consists of a bench and guiding rails system on which two tools come and facilitate the assembly of the CU. The rear part of the camera composed by the Mangin mirror, its tube support and the focal plane is decoupled from the front bell using the Camera Optics Integration Tool (COT). After sliding back, we can rotate the part to have a direct access to the focal plane, this allows to easily mount the focal plane.

When the integration is complete, one can rotate the part and move it back to fix it on the front bell using precise interfaces that minimize repositioning errors.

The cryostat is then placed on the unit station using a tool designed to easily plug the cryostat on the front bell. The tool is composed of two functional elements, a guiding system, made with two precision trolleys and rails and a mass compensation system, with an adjustable axis, a compression spring with claw for a quick compression of the springs. A set of flexible parts allows for small movements on every axis during the final positioning

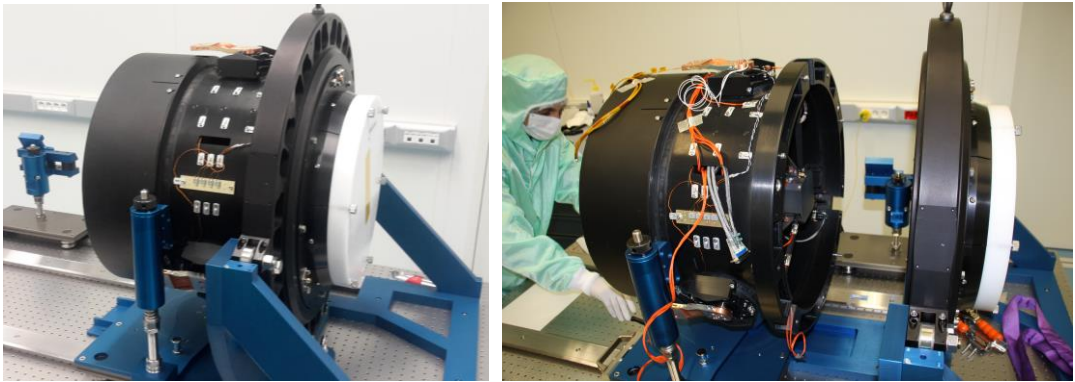

Figure 6 Assembly of the CU using the AIT Station

We have been using this station more than a dozen of time in the past seven months. The tools working well within the required performance. It really makes easy the assembly of the camera. Its usefulness could not be debated, in particular due to the unexpected thermal issues that we encounter which has increased the number of assembly/dis-assembly of the camera.

Some improvements will be done, as such a mass compensation to hold the complete cryostat body. The nominal process did not include this step so the tool was not optimized for that particular situation.

## 2.3 Performance validation tool

The verification and validation process of the spectrograph requirements needs a dummy illumination source that allow us to perform the different tests described in validation roadmap document.

The main requirements that need to be validated by tests are: the wavebands, the spectral resolution, the image quality and the ghosts. Each of these tests requires specific sources with different characteristics, such as spectral width, bandwidth. One common parameter is that we need a uniform illumination pupil with a F/2.8 beam.

The specification of the dummy cable B is described in the following Figure 7. At the top, all the wavelength specifications needed are listed. At the bottom, the fibers that need to be illuminated (how many and where on the slit head) are listed. The link between both gives the couple [wavelength/ fiber mapping] associated to the validation of a specific requirement.

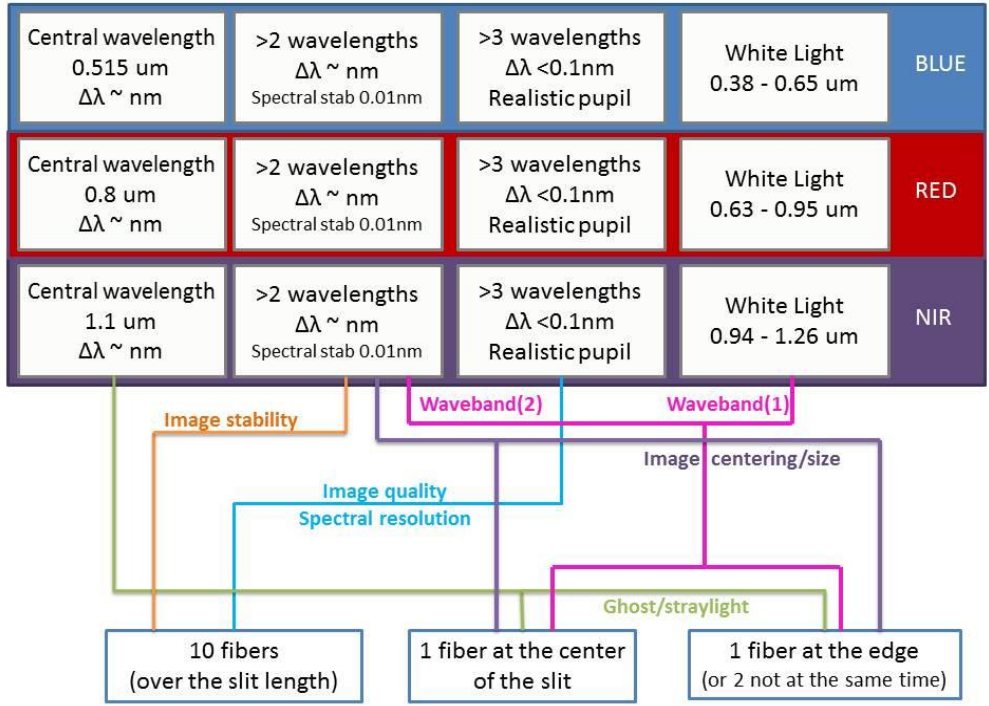

Figure 7 Illumination sources

### 2.3.1 Fiber illumination system

The PSF Dummy Cable B is based on an illumination module with a few pen-ray spectral lamps and a white light lamp linked to a double integrating sphere. This allows us to create an uniform illumination with all the required lighting characteristics.

An optical setup base on little collimators creates a realistic pupil with a F/2.8 input beam. A fiber system is made to inject the light in the Cable A. The global concept of the Dummy Cable B is presented below in Figure 8.

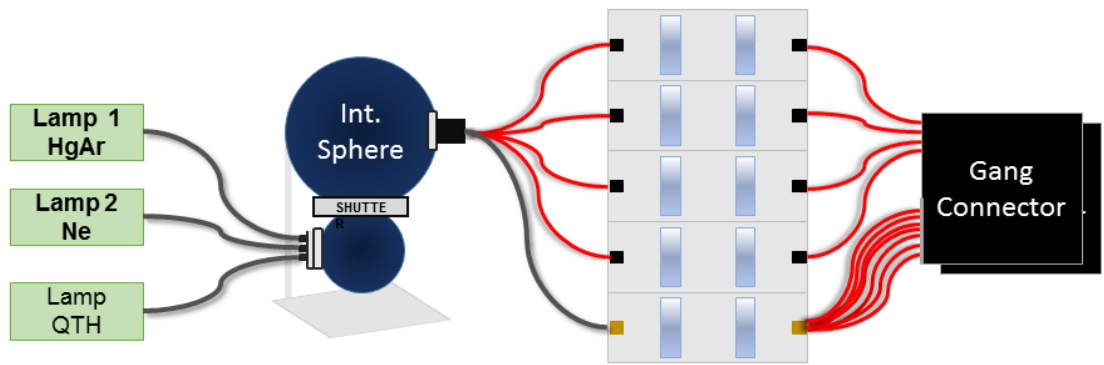

Figure 8 Dummy cable B overview

# 3. VIS CU THERMAL PERFORMANCE VALIDATION

### 3.1.1 Thermal design description

The main requirement for the visible Camera Units (VIS-CU) is to provide a detector temperature below 163K, allowing for sufficiently low dark current. Since there is no thermal background constrain, most of the optical and mechanical parts of the camera can remain warm, but in order to facilitate the thermal design and also for commonality with the near infrared camera units, the entire system is contained within a vacuum vessel to which the first corrector lens is the window. The design of the VIS-CU is shown in **Erreur ! Source du renvoi introuvable.** (left), the labelled elements indicating the cold path linking the cryo-cooler to the focal plane array. A set of thermal straps links the cryo-cooler to a triangular (tricorn-shaped) plate referred to as the thermal spreader, to the extremities of which are fixed the thermal bars, running up along the camera tube parallel to the optical axis at 120 degrees to each other. These are again thermally strapped to a set of copper rods, see **Erreur ! Source du renvoi introuvable.** (right), ensuring the thermal and mechanical interfaces between the cold path and the detector spider structure holding the detector box in the focus of the Schmidt camera. The focal plane array is inserted into the detector box in front of which is attached the field lens which is also cold.

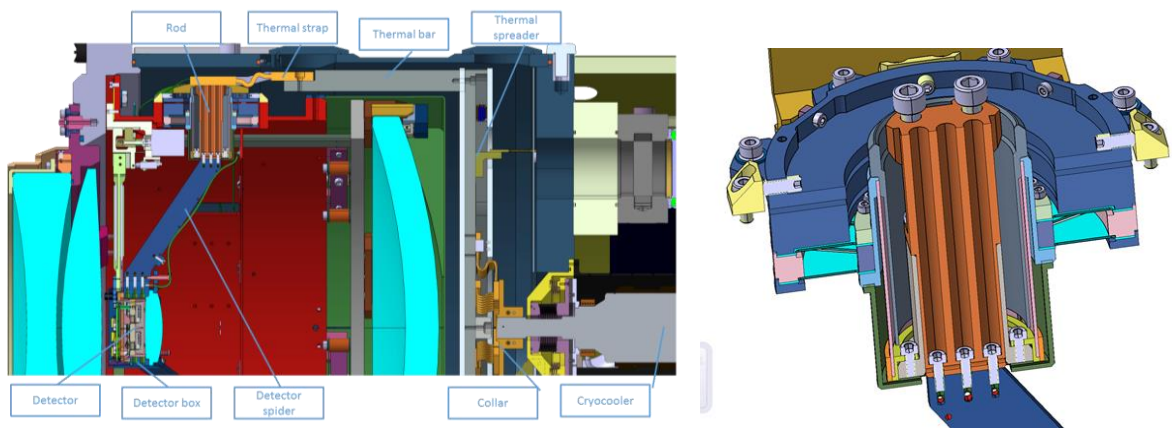

Figure 9 Visible CU thermal design

### 3.1.2 Test results

LAM is in charge of the assembly and test of the VIS-CU with the parts coming from WS and JHU. While the Dewar assembly, containing the vacuum vessel, cryo-cooler, and cold path up to the thermal strap, was tested independently at JHU before shipping to LAM, the thermo-mechanical structure holding the focal plane array, provided by Winlight, could not be independently tested. The first thermal test of the complete VIS-CU was therefore done at LAM.

It soon appeared that the thermal load was significantly higher than expected, making it impossible to reach the cryo-cooler set-point temperature (100K). The identification of the reasons for this discrepancy has proved quite complex and it is still ongoing. To improve the diagnostic capacity of our setup we have added several temperature sensors along the thermal path allowing extensive temperature mapping of the system.

Several hurdles have been faced during this work, and jumping them have allowed us to improve our system considerably. For example, we found an inconsistency between the system's temperature sensors and the additional independent temperature sensors which was tracked down to an incorrect calibration of the camera unit's built-in sensors. We also discovered that the cryo-cooler had been damaged during the transportation, leading to a reduction of cryo-cooler performance. Unfortunately, its replacement did not improve the thermal performance of the system.

Studying the thermal fluxes at different points in the design, both along the thermal path and in the warm structure, led us to identify an issue in the Rod structure, a complex and critical piece that isolates the cold path from the rest of the system, see **Erreur ! Source du renvoi introuvable.** (right). With some 15 W drained along the thermal bars and only 5 W running along the thermal spiders, nearly 10 W were seen to disappear within this structure. On the other hand, a similar amount of power was found to transit across the CuBe blades fixing the warm part of the rod structure to the surrounding mechanical structure, indicating the presence of a leak, conductive or radiative, at this point. When a simplified rod structure consisting of a simple, MLI-clad rod held in place by G10 strips was mounted in place of the nominal rod, a phenomenon that had occurred previously but not to the same extent caught our attention. While the total cooling power provided by the cryo-cooler, referred to as the "lift," was lower at the beginning of the run with the simplified rod, it increased during the run, stabilizing at a level similar to that seen with the original rod structure.

Our current thinking explains these various observations as a symptom of outgassing. The warm structure containing numerous parts made from different materials and with various types of coatings can be expected to produce a variety of molecules of which water is probably dominant but not necessarily unique. These volatiles will sooner or later be trapped, either by the various pumps contained in the Dewar system (a turbo pump and two ion pumps), or by the cold surfaces in the system. Indeed, the power drift observed during all our cold runs indicates that the latter pumping mechanism is present and could even be dominant. Furthermore, the change seen after modifying the rod structure, leading to a much improved venting of the inner camera cavity, could signify that this cavity is the main source of outgassing in the system. In the nominal system, the cold parts of the rod structure would capture the majority of the molecular pollution, rapidly leading to a catastrophic degradation of its performance, while when venting improved, the molecules would be trapped all along the cold path, leading to the stronger drift observed and the modified thermal profile.

Based on this hypothesis, tests are now ongoing where a moderate bake-out of the system is introduced between cold runs: heaters mounted on the outside of the vacuum vessel heat the system to some 35 degC, the maximum allowed for safety reasons. After two week-long baking periods we have seen a significant improvement of the thermal performance, and we are now preparing to reinstall the nominal rods for a new end-to-end test of the system. In parallel, actions for improved ventilation of the inner cavity are defined, and a procedure for efficient bake-out to apply to the remaining vacuum vessels is being prepared.

## 4. SYSTEM

### 4.1 ICD

In such large instrument with several hardware contributors, the interfaces become important. In the PFS case, the number of interface is quite high and requires a dedicated tool to manage them. The project has chosen Bugzilla, a bug tracking system to handle the interface process. Each interface is entered as an issue and the stakeholders have to attach a drawing. The discussion on the interface is done using the comments feature and drawing update can be done by replacing the old drawing that becomes obsolete. Then the issue's assignee has to create a drawing bundle and ask for an agreement. Each stakeholder responsible has to approve the interface via an approval flag, finally a super-approval is done by the ICD manager.

For the PFS spectrograph system, there are 71 ICDs, the maximum number of comments is 49 while the mean is 14.

### 4.2 Anomaly reporting

Anomaly tracking is really important especially for a system with several identical modules, in order to not reproduce issues that appears on a previous module. We also used Bugzilla to track the anomalies. The anomaly hierarchy is simply based on the product tree down to level 2 and with the module number. We also added a general and a software category. The general corresponds to bugs that are not fully diagnosed or corresponded to a non-listed component. The gen could also be high level issue that implies several units, each of the resulting bugs must be resolved before resolving general.

## 5. LOGISTICS

One challenge for the PFS Spectrograph collaborating institutes is the transport of large, costly and sensitive equipment between Japan, Brazil, France, USA (Baltimore and Hawaii). Every component of the PFS Spectrograph developed in one of the institutes needs to be shipped to LAM, then integrated, tested and validated at LAM in France. Each of the four very identical spectrograph modules (SM) will fully tested at LAM. Each of the SMs will then be boxed and shipped to Hawaii.
LAM is partnering with Ulisse, a CNRS logistics unit, specialized in transport, custom and logistics for large science instrument and programs.

As shown in Figure 1, each spectrograph module is composed of several units: Fiber Cable Assembly, Blue and Red Camera Units, Near infrared Camera Unit, Central Optical Assembly, Collimator Mirror Assembly, Optical Bench Assembly, Blue Disperser Assembly and Red Disperser Assembly as well as the electronics. For these 10 assemblies, we have designed dedicated shipping containers from a set of dedicated specifications (depending on the assembly): mechanical interface with box plate, damping (for optical components), dimensions, mass, cleanliness procedure, humidity and temperature range, lifting procedure and jigs, markings, interface for shock-log, etc. All transports are handled by trucks and planes.

PFS spectrograph containers and flycases were developed with E3Cortex (based in France). All assemblies that include optical elements are first mounted on a base plate structure inside a flycase, that is bolted on a damping system inside an waterproof and sun-resistant fiberglass-reinforced container (see pictures below).

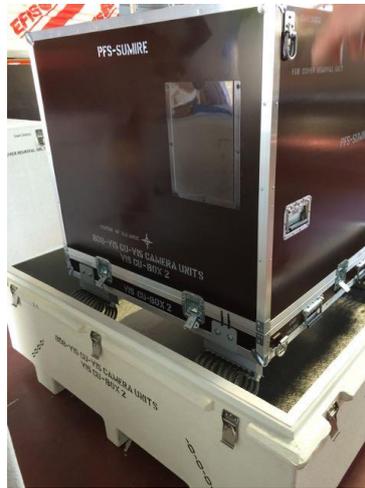

**Figure 10 : photo of one the shipping container for PFS**

As we fabricated two or three containers per type of assembly, we need to track the location of the containers throughout the transport operations. All transport operations are handled through Ulisse sevices, using their forms and following a procedure adapted for the PFS project. We also contracted a general insurance policy through Ulisse for all of our hardware during transport till the commissioning at SUBARU. Last, but not least, are the interface with French customs. We import all PFS goods (of important dollar value) temporarily in France, integrate, align and test them, then ship them elsewhere. In order to optimize the custom declaration process, minimize any cost and save time we have invited custom officers at LAM, presented the project and initiated a procedure to minimize time and cost on both sides. We share with French customs and Ulisse a tracking system where all entries and exit from France of PFS components are planned and tracked. Once declared, components can be directly shipped to LAM without any stop in Paris custom hubs. Custom officers in Marseille can visit LAM at any point and perform a review of all imported goods.

## 6. SOFTWARE

### 1.1 Architecture

Control software for complex and distributed systems is challenging.

In the instrument, each device is handled and controlled by a high level programming structure called Actor. They follow a well-defined structure where some functionalities (command handling, network protocols, etc ...) are already implemented. They are designed to encourage independent development of hardware and software subsystems, and their eventual integration. Furthermore, this factorized approach also greatly improves testing and maintenance, by making hardware more directly accessible.

These actors are connected to a central hub machine called MHS (Messaging Hub System) which runs a service called "Tron".

Tron is a relatively simple distributed communication system using twisted, an event-driven network programming framework written in Python and licensed under the MIT License.

The very basic principle, shown in Figure 11, is that tron central hub accepts commands received from *commanders* (client, actors …) and dispatches them to actors. They will reply to commands and generate status keywords, also via the hub. As a consequence, the only significant requirement is that each component or program must come with a published dictionary describing status keywords which fully define the state of the module, and that in operation those keywords are kept updated. Moreover, an archiver service track and store these keywords in a PostgreSQL database.

For example, Entrance Unit is an actor composed of several assemblies such as:

- IIS : Internal Illumination Sources
- BIA : Back-Illumination Assembly
- FSA : Fiber Slit Assembly mounted on its hexapod
- Shutter : Shutter controller
- ENU : Environment (Temperature, humidity)

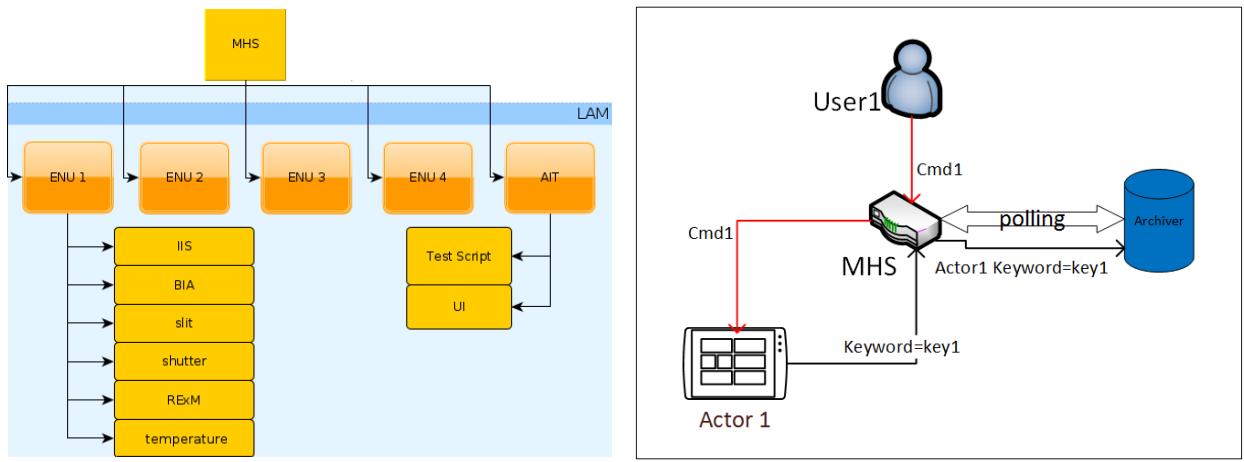

Figure 11 Left: ENU actor. Right: MHS software

## 1.2 AIT SW

For AIT activities we have several needs in terms of software. For thermal acceptance, some tools have been developed to display and monitor data from the instrument.

Lib_dataQuery is a generic library developped to answer to a common need of all tools. It simplifies interaction with the database by the use of high level programming functions.

plotData is a flexible tool which allows you to plot data and custom your figures. It' has been made to be both a real time monitoring software and data display. We use it to monitor pressure, temperature sensors …

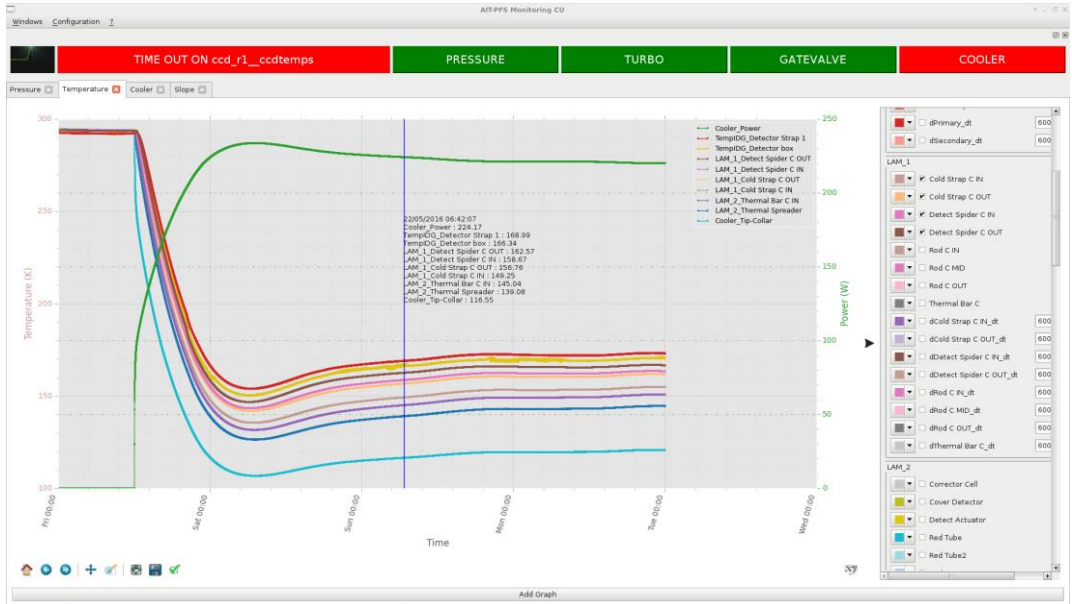

Figure 12 PlotData window

monitorData is a monitoring tools which gives a global view of all Camera unit status.

As some devices are criticals, we have added some visual alarms when they reach an unexpected state.

JabberBot is using XMPP protocol which allow users to get instrument's status and alarm from a smartphone or a web browser

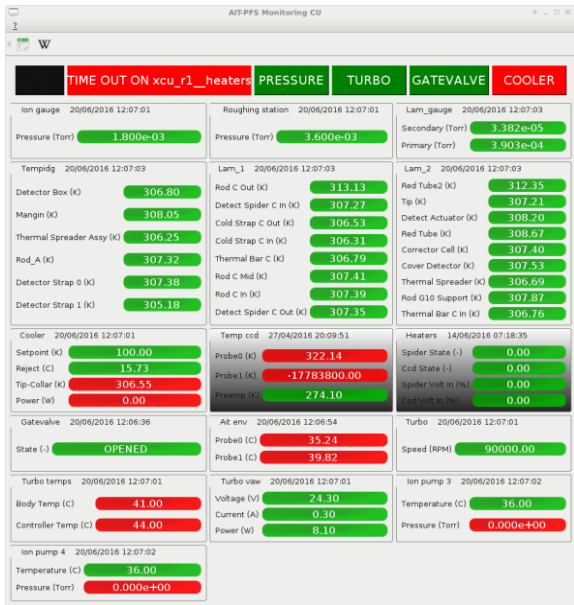 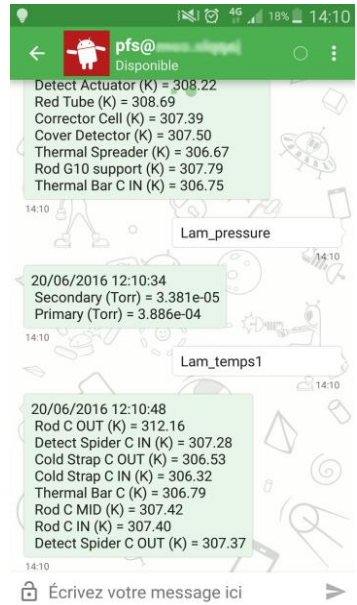

Figure 13 Left: MonitorData window. Right: Jabber bot client

## 7. CONCLUSION

We have presented the on-going activities for PFS AIT at LAM. The development of the individual assemblies for the spectrograph is progressing well, yet with delays compared to the original schedule. The optical performance for the one-channel should be completed in the fall of 2016, followed by the AIT of the spectrograph module #1. In parallel, we will have resolved our difficulties with the thermal validation (see section 3). By the fall of 2017, we expect to have shipped and integrated at least one spectrograph module at the SUBARU telescope, allowing us to proceed with on-sky engineering & commissioning.

## 8. ACKNOWLEDGEMENT

We gratefully acknowledge support from the Funding Program for World-Leading Innovative R&D on Science and Technology (FIRST) program "Subaru Measurements of Images and Redshifts (SuMIRe)", CSTP, Japan.

## REFERENCE LINKING